\documentclass[pra,aps,showpacs,superscriptaddress,twocolumn]{revtex4-2}
\usepackage{graphicx,amsmath,amssymb,color, braket, comment}
\usepackage{xcolor}
\usepackage[colorlinks, citecolor={blue!80!black}, urlcolor={blue!80!black}, linkcolor={red!50!black}]{hyperref}

\begin{document}

\title{Vacuum-induced atomic grating}

\author{Muhammad Shuraim}
\affiliation{Department of Physics and Applied Mathematics, Pakistan Institute of Engineering and Applied Sciences (PIEAS), Nilore $45650$, Islamabad, Pakistan.}
\author{Muhammad Waseem}
\affiliation{Department of Physics and Applied Mathematics, Pakistan Institute of Engineering and Applied Sciences (PIEAS), Nilore $45650$, Islamabad, Pakistan.}
\affiliation{Center for Mathematical Sciences, PIEAS, Nilore, Islamabad $45650$, Pakistan.}
\author{Shahid Qamar}
\affiliation{Department of Physics and Applied Mathematics, Pakistan Institute of Engineering and Applied Sciences (PIEAS), Nilore $45650$, Islamabad, Pakistan.}
\affiliation{Center for Mathematical Sciences, PIEAS, Nilore, Islamabad $45650$, Pakistan.}
\author{Muhammad Irfan}
\email[Corresponding author: ]{m.irfanphy@gmail.com}
\affiliation{Department of Physics and Applied Mathematics, Pakistan Institute of Engineering and Applied Sciences (PIEAS), Nilore $45650$, Islamabad, Pakistan.}
\affiliation{Center for Mathematical Sciences, PIEAS, Nilore, Islamabad $45650$, Pakistan.}
\date{\today}

\begin{abstract}

Atom-field interactions, induced by the vacuum of the electromagnetic field, exhibit a variety of fundamental phenomena and effects.
In this paper, we study the electromagnetically induced atomic grating due to the vacuum state of the radiation field.
Using an ensemble of cold atoms, strongly coupled to an optical cavity, we show that a probe field, propagating through the atomic medium, diffracts to zeroth and first-order diffraction peaks with few photons and even by the electromagnetic vacuum field of the cavity mode.
As the number of photons in the cavity increases, the intensity of the first-order diffraction peak initially rises and then exhibits a decreasing trend.
Furthermore, we observe that the first-order peak intensity reaches its maximum at resonance for both the vacuum and single-photon cavity state.
However, as the number of photons increases further, this peak at resonance transforms into a dip, accompanied by two side peaks at off-resonance positions.
This transition from a peak to a dip may potentially be used to distinguish the quantum state of the cavity.

\end{abstract}

\maketitle
\newpage
\section{Introduction}
The quantum coherent effects in light–atom interaction are vital for manipulating the optical properties of a medium.
These effects play a significant role in studies of optical devices based on the atomic medium, quantum computing~\cite{pellizzari_decoherence_1995, turchette_measurement_1995, rauschenbeutel_coherent_1999}, and advancing nonlinear effects in quantum optics~\cite{firstenberg_nonlinear_2016}.
A notable example is electromagnetically induced transparency (EIT)~\cite{harris_electromagnetically_1997,boller_observation_1991,harris_nonlinear_1990,fleischhauer_electromagnetically_2005} in atomic ensembles that provide an impressive degree of coherent control of the optical properties of the medium~\cite{boller_observation_1991,fleischhauer_electromagnetically_2005,kash_ultraslow_1999,finkelstein_practical_2023}.
In EIT, the atomic medium becomes transparent to a weak probe field within a narrow frequency range in the presence of a strong control laser field.

In the past, the effects of quantum coherence induced by the vacuum of the electromagnetic field have been extensively studied both experimentally and theoretically.
Such light-atom interactions, induced by the vacuum of the electromagnetic field, exhibit purely quantum phenomena such as vacuum Rabi splitting~\cite{field_vacuum-rabi-splitting-induced_1993, schuster_resolving_2007, agarwal_vacuum-field_1985, miller_trapped_2005,PhysRevLett.110.066802}, vacuum-induced transparency~\cite{tanji-suzuki_vacuum-induced_2011,guo_vacuum_2015}, vacuum-induced coherence~\cite{hou_effects_2006}, vacuum-induced modification of four wave mixing~\cite{chen_vacuum-induced_2015}, vacuum-induced memory~\cite{qin_vacuum-induced_2018, fan_vacuum_2018}, 
Lamb shift \cite{lamb_fine_1947}, spontaneous emission~\cite{gea-banacloche_vacuum_1988}, and so on.
In the context of EIT, it has been shown that cavity-based EIT strongly depends on the quantum state of the cavity field~\cite{gor_photon_number_2010}.
It has been experimentally demonstrated that the transparency window of EIT still appears even when there are no photons in the control field, within a cavity interacting with three-level atoms in strong coupling regime~\cite{tanji-suzuki_vacuum-induced_2011,field_vacuum-rabi-splitting-induced_1993}.

Remarkably, in an EIT system, if we replace the strong control field with a standing field, we obtain spatial modulation of the amplitude and phase of the probe field.
As a result, the atomic medium acts as a tuneable atomic grating which diffracts the incident probe light into zeroth and higher-order diffraction peaks, similar to traditional optical gratings.
This phenomenon, known as electromagnetically induced grating (EIG) was first proposed by Ling et al.,~\cite{ling_electromagnetically_1998} and observed by Mitsunaga et al. in sodium atoms~\cite{mitsunaga_observation_1999}.
Since then EIG has been widely explored in a variety of different systems~\cite{yuan_observation_2019, tabosa_transient_1999, asghar_electromagnetically_2016, zhang_observation_2020, badshah_electromagnetically_2023, Zhao2018, Shui2020, PhysRevA.105.063511} with many promising potential applications~\cite{brown_all-optical_2005, Zhao_alloptical-2010, ma_rydberg_2019, cardoso_electromagnetically_2002, wen_talbot_2011, anees_talbot_2022, asghar_electromagnetically_2019}.
Most EIG studies follow a semiclassical approach, where the control field is considered to be a strong classical standing field.
Motivated by the vacuum-induced quantum coherence effects mentioned earlier, we explore the effects of the vacuum state of the electromagnetic radiations in EIG, which has not been reported so far to the best of our knowledge.

In this paper, we investigate an ensemble of three-level $\Lambda$-type atoms within a single-mode cavity, where the cavity field acts as the quantized control field.
We consider a strong atom-field coupling regime such that the coupling coefficient is equal to or greater than the excited state's spontaneous decay rate and the lifetime of the cavity photons~\cite{tanji-suzuki_vacuum-induced_2011}.
Our findings reveal that a vacuum state of the cavity field diffracts incident probe light into zeroth- and first-order diffraction peaks, resulting in the formation of a vacuum-induced atomic grating (VIAG).
As the number of photons in the cavity increases, the intensity of the first-order diffraction peak initially increases and then follows a decreasing trend.
Interestingly, the first-order peak intensity reaches its maximum at resonance for both the vacuum state and the single-photon Fock state. 
However, as the photon numbers continue to increase, the resonance peak splits and transforms into a dip with two side peaks occurring at off-resonance positions.
Our results show that with a sufficiently large number of photons, the traditional EIG results are recovered.
This work offers a new approach to coherently manipulate the intensity of EIG  with a vacuum or few photons state. 

\section{Theoretical Model and System Hamiltonian}
\begin{figure}
    \centering
    \includegraphics[width = 0.475\textwidth]{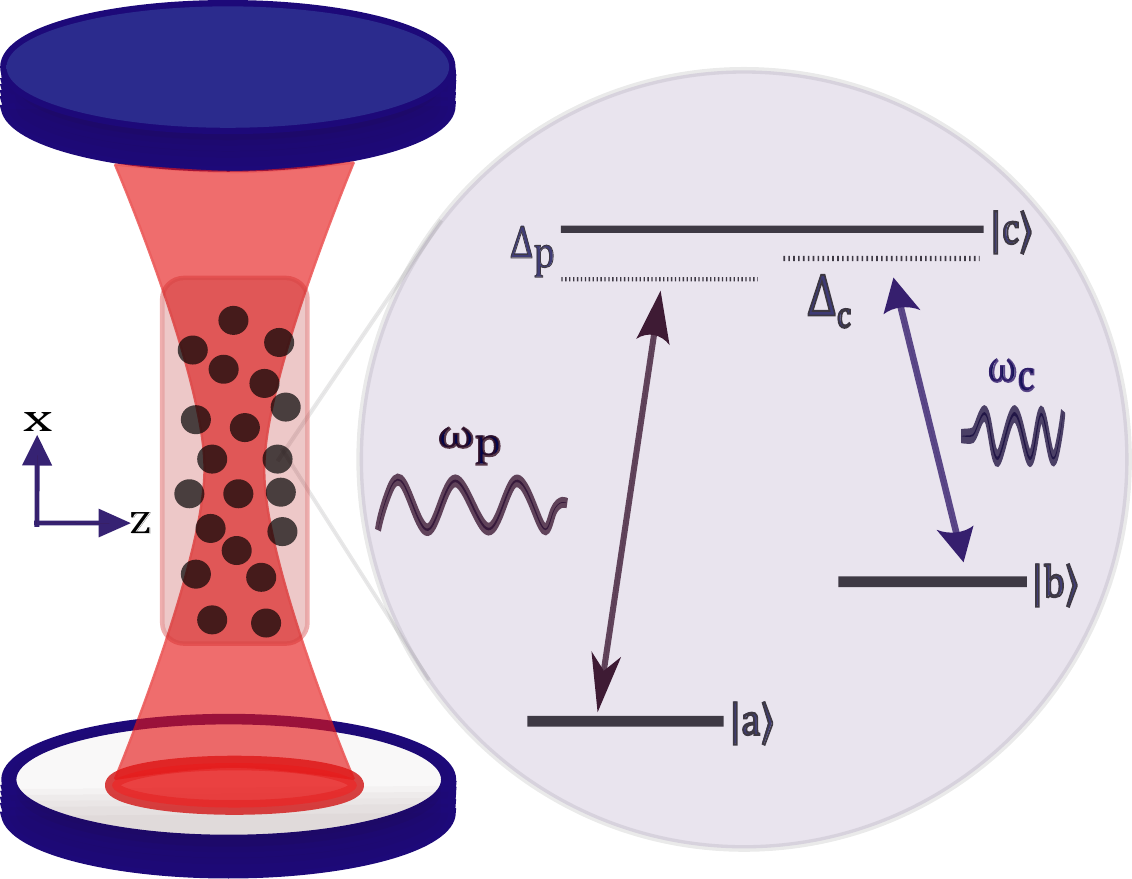}
    \caption{Schematics of the system under consideration with an ensemble of three-level atoms in $\Lambda-$configuration inside a single-mode cavity. The cavity field having frequency $\omega_c$ interacts with atoms via $\ket{c} \leftrightarrow \ket{b}$ transition. A probe field of frequency $\omega_p$ is applied which interacts with $\ket{c} \leftrightarrow \ket{a}$ transition.}
    \label{fig:system-viag}
\end{figure}

We consider an ensemble of $\Lambda$-type three-level atoms placed inside a single-mode cavity, as shown in Fig. \ref{fig:system-viag}.
An incident probe field of frequency $\omega_p$ drives the atomic transition  $\ket{a}-\ket{c}$, whereas the cavity mode of frequency $\omega_c$ interacts with atoms via transition $\ket{b}-\ket{c}$.
The Hamiltonian of the system under the rotating-wave approximation, assuming no dissipation terms, is written as follows:
\begin{equation}\label{eq:Hamilt}
    H = H_{0} + H_{I}.
\end{equation}
Here, the unperturbed part is
\begin{equation}
    H_{0} = \hbar(\omega_{a}\ket{a}\bra{a}+\omega_{b}\ket{b}\bra{b} + \omega_{c^{'}} \ket{c}\bra{c}),
\end{equation}
with $\omega_a$, $\omega_b$, and $\omega_{c^{'}}$ are the frequencies of the levels $\ket{a}$, $\ket{b}$, and $\ket{c}$, respectively such that  $\omega_{cb}=\omega_{c^{'}}-\omega_b$ is the frequency of $\ket{c}-\ket{b}$ transition and $\omega_{ca}=\omega_{c^{'}}-\omega_a$ represents the frequency of $\ket{c}-\ket{a}$  transition.
The interaction part of the Hamiltonian is

\begin{equation}\label{eq:Hi}
    H_{I} = -\frac{\hbar}{2}(\Omega_{p} e^{-i\omega_p t}\ket{c}\bra{a} + \Omega_{c} e^{-i\omega_c t}\ket{c}\bra{b} + H.c.).
\end{equation}
Here, $\Omega_{p}$ is the Rabi frequency of the probe field.
The position-dependent Rabi frequency of the cavity field is 
$\Omega_{c}= \Omega_{0}\sin(\frac{\pi x}{\Lambda})$.
The parameter $\Lambda$ is the spatial period of the cavity field.
The amplitude of the cavity field's Rabi frequency $\Omega_0$ is described as follows \cite{tanji-suzuki_vacuum-induced_2011}:
\begin{equation}\label{eq:quantized}
    \Omega_{0} = 2g\sqrt{n_c + 1},
\end{equation}
with $g$ the atom-field coupling constant and $n_c$ the number of photons in the cavity mode.
It is important to note that although the Hamiltonian in Eq.~(\ref{eq:Hi}) is written in semiclassical form for simplicity, we quantized the cavity field in Eq.~(\ref{eq:quantized}).
If the cavity mode is in a vacuum state, the Rabi frequency $\Omega_0=2g$, which can be large if the system is in a strong coupling regime.
It is shown that if the coupling strength $2g$ is larger than the decoherence rate $\gamma_{ab}$ of energy levels $\ket{a}$ and $\ket{b}$, the system exhibits vacuum-induced transparency i.e., a dip in the absorption spectrum similar to the electromagnetic-induced transparency~\cite{tanji-suzuki_vacuum-induced_2011}.
When atoms are placed inside the cavity, the quantized cavity field interacts with atoms via transition $\ket{b}-\ket{c}$.
The coupling strength $g$, can be described in terms of single-atom cooperation parameter $\beta$ define as:
\begin{equation}
    \beta = \frac{4g^2}{\Gamma_{ca}\kappa}.
\end{equation}
Here, $\Gamma_{ca}$ is the spontaneous emission rate from state $\ket{c}$ to $\ket{a}$ and $\kappa$ is the cavity decay rate.
For strong coupling limit, $\beta$ exceeds unity~\cite{tanji-suzuki_vacuum-induced_2011}. 

The optical response of the medium to the incoming probe field is determined by the complex susceptibility $\chi$ which directly relates to the density matrix element $\rho_{ca}$.
The density matrix element $\rho_{ca}$ is obtained by solving density matrix equations by incorporating the dissipation terms~\cite{cohentannoudji_atomphoton_1998, fleischhauer_electromagnetically_2005}.
As a result, the complex susceptibility is given by~\cite{tanji-suzuki_vacuum-induced_2011} (see the Appendix for details):

\begin{widetext}
\onecolumngrid
\begin{equation}
\label{eq:chi}
\chi=N_0 \cdot \left[\frac{4 \delta \left(-\left|\Omega_{c}\right|^2+4 \delta \Delta_p\right)+4 \Delta_p \gamma_{ba}^2+i 8 \delta^2 \gamma_{ca}+i 2 \gamma_{ba}\left(\left|\Omega_{c}\right|^2+\gamma_{ba} \gamma_{ca}\right)}{\left|\left|\Omega_{c}\right|^2/\gamma_{ca}+\gamma_{ba}-4 \Delta_p \delta/\gamma_{ca}+i 2 \delta+i 2 \Delta_p \gamma_{ba}/\gamma_{ca}\right|^2}\right]\frac{1}{\gamma_{ca}},
\end{equation}
\end{widetext}
\twocolumngrid
\noindent
where $N_0 = \frac{\left.\left|\mu_{ca}\right|^2 N\right.}{\hbar\varepsilon_0 \gamma_{ca}}$, $N$ is the number of atoms per unit volume, $\Delta_p = \omega_{ca} - \omega_p$ is the probe field detuning, $\Delta_c = \omega_{cb} - \omega_c$ is the control field detuning, and $\delta = \Delta_p - \Delta_c$ is the two-photon detuning.
The decoherence rates of energy levels $\ket{c}$ and $\ket{a}$ is $\gamma_{ca} = \gamma_c + \Gamma_{cb} + \Gamma_{ca}$ while for energy levels $\ket{a}$ and $\ket{b}$ is $\gamma_{ba} = \gamma_b$.
Here, $\gamma_b$ and $\gamma_c$ are the dephasing rates of the energy levels $\ket{b}$ and $\ket{c}$, respectively, whereas $\Gamma_{cb}$ and $\Gamma_{ca}$ are the decay or the spontaneous emission rates from the state $\ket{c}$ to the state $\ket{b}$, and $\ket{c}$ to the state $\ket{a}$, respectively.
The expression for $\chi$ can be written in terms of the cooperativity parameter $\beta$ by considering $\Gamma_{ca} \approx \gamma_{ca}$~\cite{tanji-suzuki_vacuum-induced_2011}.

Since we are interested in studying vacuum-induced atomic grating, we first describe the mathematical formalism of atomic grating.
The output amplitude of the probe field having wave vector $k_p$ propagating through the three-level medium of length $\ell$ is given by: 
\begin{align}\label{Eq: amplitude}
    E_p(x, z=\ell) &= E_p(x, 0) e^{i k_p\chi \ell/2},\\ \nonumber
    &= E_p(x, 0) e^{-k_p\chi^{''} \ell/2} e^{i k_p\chi^{'} \ell/2}.
\end{align}
Here, $E_p(x, 0)$ is the probe-field amplitude (assumed to be a plane-wave) at $z=0$, whereas $\chi^{''}$ and $\chi^{'}$ are the imaginary and real parts of the complex susceptibility, respectively.
The real part of susceptibility shows the dispersion, while its imaginary part characterizes the absorption of the probe field.
Therefore, the transmission function for the probe field is given by:
\begin{equation}\label{Eq: fraunhofer diff}  
    T(x, \ell) = e^{ik_p \chi\ell/2}.
\end{equation}
The transmission function characterizes the variation of the output probe field amplitude after propagation through the medium.

The Fraunhofer diffraction is proportional to the Fourier transformation of the product of amplitude of probe field $E_p(x, 0)$ and the transmission function $T(x, \ell)$ which is given as follows~\cite{ling_electromagnetically_1998}:
\begin{equation}
    F_p(\theta) = C \int_{-\infty}^{+\infty} E_p(x, 0) \cdot T(x, \ell)e^{-2\pi i\cdot \Lambda x \sin(\theta)/\lambda_p} dx,
\end{equation}
with C the proportionality constant, $\Lambda$ the space period in the $x$-direction, and $\theta$ is the angle of diffraction with the $z$-direction.
The intensity distribution profile of the grating phenomenon for a single spatial period, considering the assumptions of Fraunhofer diffraction and Kirchhoff's diffraction theory \cite{ling_electromagnetically_1998}, is provided as follows:
\begin{equation}
    I_p(\theta) = |F_p^n(\theta)|^2 \cdot \frac{\sin^2(M\Lambda\pi\sin(\theta)/\lambda_p)}{\Bigl(M\sin(\Lambda\pi\sin(\theta)/\lambda_p)\Bigr)^2},
\end{equation}
where,
\begin{equation}\label{Eq: intensity}
    F_p^n(\theta) = \frac{1}{\Lambda}\int_0^\Lambda T(x, \ell)e^{-2\pi i\cdot x \sin(\theta)/\lambda_p} dx.
\end{equation}
Here, $M$ represents the number of space periods in VIAG, $\lambda_p$ is the wavelength of the classical probe field.
Since our primary focus is on first-order diffraction, we calculate the intensity profile $I_p(\theta)$ over the first-order diffraction angle.
This angle is determined from the Bragg diffraction equation $\sin(\theta_n) = n\lambda_p/\Lambda$, where $n = 1$ for first-order diffraction.
The intensity expression for first-order diffraction, $I_p(\theta_1)$, is provided as follows:
\begin{equation}
    I_p(\theta_1) = \left| \frac{1}{\Lambda} \int_0^\Lambda T(x, \ell)e^{-2\pi i \frac{x}{\Lambda}}\right|^2 dx.
\end{equation}
\begin{figure}[t]
\includegraphics[width=\linewidth]{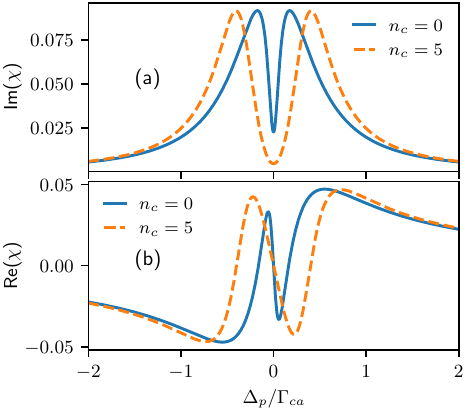}
\caption{Response of probe field: (a) Imaginary part of $\chi$ and (b) Real part of $\chi$ as a function of $\Delta_p$ for $n_c=0$ and $n_c=5$, assuming that the coupling field is resonant with the atomic transition. The parameters used in calculations are given in the text, considering $x= \Lambda/2$.}
\label{fig:chi}
\end{figure}
\begin{figure}[t]
\includegraphics[width=\linewidth]{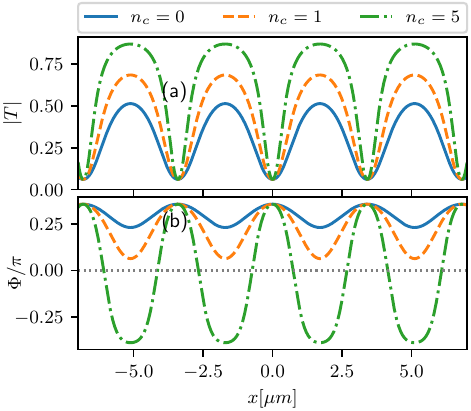}
\caption{(a) Periodic amplitude modulation (b) Phase modulation of the transmission function as a function of spatial dependence $x$ for different choices of $n_c$. For the top panel, detuning in the probe and coupling field is considered to be zero while in the bottom panel, $\Delta_p = 0.25\Gamma_{ca}$ as the real part of $\chi$ becomes zero when probe detuning is zero, as shown with the dotted gray line.}
\label{fig:transmission_x}
\end{figure}
\section{Results and Discussions}
In this section, we discuss the results of our numerical simulations, first starting with the analysis of complex susceptibility.
In Fig.~\ref{fig:chi}, we plot the real and imaginary parts of the susceptibility as a function of probe-field detuning ($\Delta_p$) for two different cases of photon numbers in the cavity mode.
For the numerical simulations, we adopted most of the parameters from the experimental realization of VIT by Tanji-Suzuki et al., \cite{tanji-suzuki_vacuum-induced_2011}, where they used Cesium atoms as three-level atoms.
The energy levels of our interest are $\ket{a} \equiv \ket{6S_{1/2}, F= 3, m_F= 3}, \ket{b} \equiv \ket{6S_{1/2}, F= 4, m_F= 4}$, and $\ket{c} \equiv \ket{6P_{3/2}, F= 4, m_F= 4}$.
We used the following parameters for our numerical simulations: $\lambda_p = 852~$nm, $\mu_{ca} = 3.79 \times 10^{-29}~$Cm, $\epsilon_o = 8.85 \times 10^{-12}~$Fm$^{-1}$, $N = 10^{12} ~$cm$^{-3}$, $\beta = 3.2$, $\Gamma_{ca} = 2\pi \times 5.2 \times 10^{6}~$Hz, $\kappa = 2\pi \times 173 \times 10^{3}~$Hz, $\ell = 8.0~\mu$m, $\Lambda = 4.0 \lambda_p$, $k_p = \frac{2\pi}{\lambda_p}$, $M = 5$, $\gamma_{ba}=\kappa$~\cite{fan_vacuum_2018}.
Fig.~\ref{fig:chi}(a) shows the variation of the imaginary part of linear susceptibility (Im$[\chi]$) with the probe detuning ($\Delta_p$).
The solid curve represents the case when there are no photons ($n_c=0$) in the cavity and the strength of the Rabi frequency of the coupling field is solely determined by the atom-field coupling strength $g$, i.e., $\Omega_0=2g$. 
We note that there is a dip in the absorption spectrum of the probe field at resonance ($\Delta_p=0$) with two side peaks, an indication of a vacuum-induced transparency window~\cite{tanji-suzuki_vacuum-induced_2011}.
It must be mentioned that in the typical semiclassical treatment of electromagnetic-induced transparency, we get a single absorption peak at resonance when the control field is absent ($\Omega_{c}=0$).
The linewidth of the absorption peak is estimated as proportional to the decay rate $\gamma_c$.
The dashed curve in Fig.~\ref{fig:chi}(a) shows that for $n_c=5$, the minimum of the dip gets deeper as compared to the $n_c=0$ case along with the increase in width of the transparency window.
Therefore, by increasing the number of photons the dip amplitude approaches zero absorption.
Fig.~\ref{fig:chi}(b) shows the plot of Re$[\chi]$ as a function of probe-field detuning for $n_c=0$ (solid cure) and $n_c=5$ (dashed curve).
At resonance ($\Delta_p = 0$), the magnitude of the Re$[\chi]$ is zero, however for the off-resonant case, the phase modulation strongly depends on the number of photons in the cavity.
\begin{figure}[t]
\includegraphics[width=\linewidth]{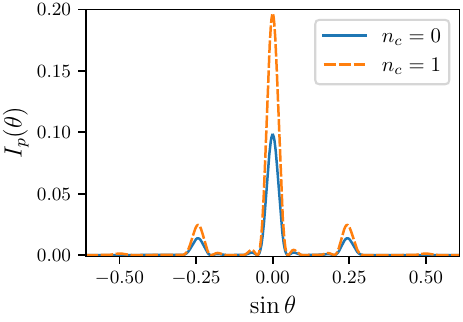}
\caption{Intensity profile of the Fraunhofer diffraction as a function of $\sin{\theta}$ when both the probe and coupling detunings are considered to be zero. Here, $\Lambda/\lambda_p = 4$ and the rest of the parameters are given in the text.} 
\label{fig:I_pvstheta}
\end{figure}

We are interested in studying the diffraction of the probe field.
Before investigating the Fraunhofer diffraction intensity, we plot, in Fig.~\ref{fig:transmission_x}(a), the spatial dependence of the amplitude of the transmission function for three different choices of the cavity photon-number $n_c$.
The solid curve represents the case when there are no photons in the cavity, and shows that the transmission amplitude periodically varies with the spatial period ($\Lambda$) of the Rabi frequency.
We considered the resonant condition while plotting the amplitude of the transmission function.
This spatial periodic variation of the transmission function shows the formation of vacuum-induced atomic grating.
Increasing the number of photons in the cavity increases the amplitude of the transmission function.
Fig.~\ref{fig:transmission_x} (b) shows the periodic variation of the phase of the transmission function $\Phi = (k_p\ell \cdot \rm{Re}[\chi]/2)$.
Since, Re$[\chi]$ vanishes at resonance, therefore, we consider an off-resonant case ($\Delta_p=0.25\Gamma_{ca}$), which clearly shows the photon-dependent increase in phase modulation.

\begin{figure}[h]
\includegraphics[width=\linewidth]{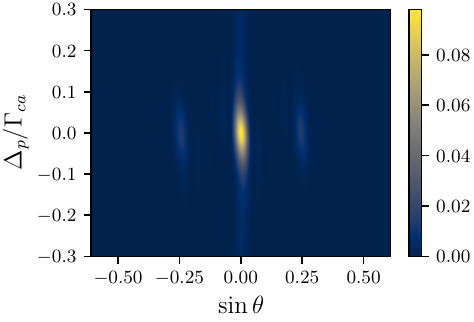}
\caption{Intensity profile of the Fraunhofer diffraction as a function of \( \sin\theta \) and probe detuning for $n_c=0$. The rest of the parameters are the same as in Fig.~\ref{fig:I_pvstheta}.}\label{fig:fig7}
\end{figure}

\begin{figure}[t]
\includegraphics[width=\linewidth]{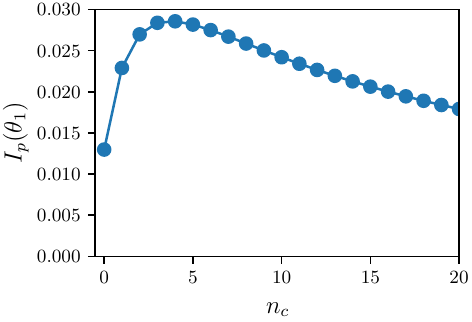}
\caption{Intensity of the first-order diffraction as a function of the number of photons available in the cavity. Here, $\sin{\theta} = \lambda_p/\Lambda$ while the rest of the parameters are the same as in Fig.~\ref{fig:I_pvstheta}.}
\label{fig:I_p1vsnc}
\end{figure}

Next, we present the intensity profile of the Fraunhofer diffraction as a function of the diffraction angle in terms of $\sin \theta$ for two different choices of $n_c$ in Fig.~\ref{fig:I_pvstheta}.
The solid curve is for the cavity vacuum state, which clearly shows the formation of vacuum-induced atomic grating with zeroth and first-order diffraction peaks.
The dashed curve shows the diffraction intensity for $n_c=1$.
The peak intensity of both zeroth-order and first-order diffraction peaks increases for the one-photon state as compared to the zero-photon case.
This indicates that the intensity of the first-order peak may be increased for a higher number of photons.
It is important to mention that the traditional classical treatment shows that there is no diffraction when the Rabi frequency of the control field is zero.
However, this is not true if we consider the quantized nature of the field, an effect reminiscent of semiclassical vs quantum mechanical treatment of a two-level system interacting with an electromagnetic field where the first does not describe spontaneous emission while the second does.
In Fig.~\ref{fig:fig7}, we plot the intensity as a function of \( \sin\theta \) and probe detuning for $n_c=0$.
It shows that the first-order peak is quite narrow as a function of the probe field detuning.
The effects of probe field detuning on first-order peak intensities for different choices of cavity photon number states are presented in the following discussion.

\begin{figure}[t]
\includegraphics[width=0.475\textwidth]{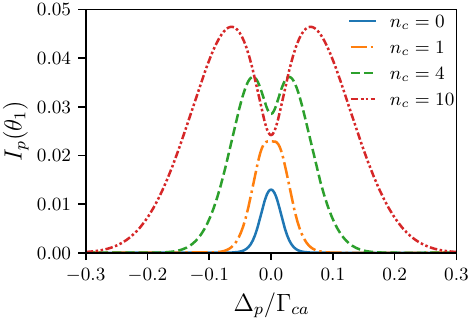}
\caption{Intensity of the first order diffraction as a function of the probe detuning. The parameters are the same as in Fig.~\ref{fig:I_p1vsnc}.}
\label{fig:I_p1vspdet}
\end{figure}

\begin{figure}[t]
\includegraphics[width=0.475\textwidth]{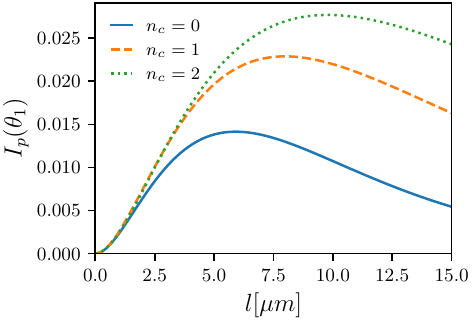}
\caption{Intensity of the first order diffraction as a function of the medium length, assuming resonant conditions. The parameters are the same as in Fig.~\ref{fig:I_p1vsnc}.}
\label{fig:I_p1vsd}
\end{figure}

In Fig. \ref{fig:I_p1vsnc}, we plot the first-order peak intensity ($\sin{\theta} \sim 0.25$) as a function of the number of photons in the cavity mode.
It shows that indeed the peak value increases initially with increasing photon numbers from zero, reaches a maximum around $n_c=4$, and then slowly decreases as a function of photon number.
The difference between the VIAG and traditional EIG discussed above is more clearly visible from Fig. \ref{fig:I_p1vsnc}.
The classical case of Ref.~\cite{ling_electromagnetically_1998} shows that the plot starts from zero intensity as a function of the control-field Rabi frequency while the quantized version shows that there is
a significant first-order peak intensity even for a vacuum state, which is quite interesting.
To understand the trend shown in Fig. \ref{fig:I_p1vsnc}, next we plot the first-order peak intensity as a function of probe-detuning for different values of cavity photon-number $n_c$ in Fig~\ref{fig:I_p1vspdet}.
It is important to note that Fig.~\ref{fig:I_pvstheta} and Fig.~\ref{fig:I_p1vsnc} are plotted for the resonant case $\Delta_p=0$.
It is therefore clear from Fig~\ref{fig:I_p1vspdet} that the peak intensity is maximum at resonance for $n_c=0$ and $n_c=1$, whereas, this peak turns to a dip at resonance for $n_c>1$.
For $n_c=4$ and $n_c=10$, our results show that the dip goes deeper as the photon number increases from $n_c=4$.
The results are in agreement with the trend previously shown in Fig.~\ref{fig:I_p1vsnc}.
It is worth mentioning that for a sufficiently large number of photons, the trend qualitatively resembles the classical electromagnetically induced grating initially discussed in Ref.~\cite{ling_electromagnetically_1998}.
This transition of a peak to a dip at resonance may have an important application in distinguishing a zero or few-photon cavity state from a larger photon number state.
Since at resonance, there is no phase modulation, therefore the grating is purely an amplitude grating.
We note that the resonance condition is optimum for zero or one photon case to obtain better first-order peak amplitude.
However, for a larger number of photons, a detuned probe field gives better first-order peak intensities due to the phase grating phenomenon where phase modulation plays an important role.
Finally, we present the effects of atomic medium length on the first-order peak intensity in Fig.~\ref{fig:I_p1vsd}.
For the given set of parameters, the intensity of the first-order diffraction reaches its maximum for $n_c=0$, when the atomic medium length is maintained between 5 µm and 6 µm.
However, for other choices of cavity photon states, the optimum length increases~\cite{Saddique_2024}.

We have used the parameters in our numerical simulations from the seminal experimental work by Tanji-Suzuki et al.~\cite{tanji-suzuki_vacuum-induced_2011}, where they demonstrated the vacuum-induced transparency in an ensemble of Cesium atoms trapped inside an optical cavity.
This experimental setup may be used with necessary modifications to observe the diffracted probe field discussed in this work.
In our numerical simulations, we chose the cooperativity parameter $\beta=3.2$, however, the recent advancements promise a much higher cooperativity factor~\cite{PhysRevLett.120.093601}.
An optimized cooperativity factor along with other optimized parameters will enhance the diffraction intensities.

\section{Conclusion}
To conclude, we proposed and analyzed vacuum-induced atomic grating in $\Lambda$-type three-level atoms inside a single-mode optical cavity.
We show that the vacuum cavity field diffracts the incident probe field to the zeroth and first-order diffraction spectrum.
We also study the effects of photon number increment on the diffraction grating, which clearly shows that this discrete photon-number-dependent grating, for sufficiently large photons, approaches the traditionally studied electromagnetically induced grating, where the control field is treated classically.
Furthermore, we found that the first-order diffraction peak intensity exhibits an important feature that if the cavity field is in zero or one photon state, the intensity is maximum at resonance.
However, when the number of photons increases further, the peak turns to a dip.
This may potentially be used to distinguish a vacuum or a few photon states from a multiphoton state.
Our findings highlight the significant role of the quantum state of the control field in shaping the diffraction patterns in a strongly coupled light-atom interaction system. Understanding these dynamics can advance the development of quantum optical devices and enhance the control over optical properties in quantum information processing and nonlinear optics.

\appendix
\section{Derivation of the linearized susceptibility}
Here, we present the derivation of the susceptibility given in Eq.~\ref{eq:chi}.
We transform the Hamiltonian (Eq.~\ref{eq:Hamilt}) under the following unitary transformation:

\begin{align}
U =  e^{i[\omega_a \ket{a}\bra{a} + (\omega_a+ \omega_p-\omega_c) \ket{b}\bra{b} + (\omega_a-\omega_p) \ket{c}\bra{c}]t}.
\end{align}
The resulting transformed Hamiltonian is given by:
\begin{align}
H &=  \hbar (\delta \ket{b}\bra{b} + \Delta_{p} \ket{c}\bra{c}) \notag \\
&\quad - \frac{\hbar}{2} \left( \Omega_p\ket{c}\bra{a} + \Omega_c \ket{c}\bra{b} + H.c. \right).
\end{align}

To obtain the dynamics of the system, we use the Von-Neuman equation.
\begin{align}
\dot{\rho} = &\ \frac{1}{i \hbar}\left[H, \rho\right] - \mathcal{L}
(\rho),
\label{eq:von}
\end{align}
with $\dot{\rho}$, the time derivative of the density matrix and $\mathcal{L}(\rho)$, the Lindblad superoperator, expressed as:  
$$
\mathcal{L}(\rho) = \sum_{n} \frac{\gamma_n}{2} \left( \sigma_n^\dagger \sigma_n \rho + \rho \sigma_n^\dagger \sigma_n - 2 \sigma_n \rho \sigma_n^\dagger \right),
$$
where $\gamma_n$ describes the decay rates, including dephasing or spontaneous emission, and $\sigma_n$ are the corresponding atomic operators.

Assuming a weak probe field, we are interested in the linear response of the system, described by the susceptibility $\chi$ which depends on the density matrix element $\rho_{ca}$ as $\chi^{(1)} = \frac{2\left|\mu_{ca}\right|^2 N}{\epsilon_0 \hbar \Omega_p} \rho_{ca}$.
To calculate $\rho_{ca}$, we solve the density matrix equations (Eq.~\ref{eq:von}), treating the probe field only up to the first order, while keeping all orders for the control field.
Under this condition, we write the following density matrix equations, assuming the Rabi frequencies to be real-valued:

\begin{align}
\dot{\rho}_{ca} &= \frac{1}{2}[-i\Omega_p \rho_{cc} + i\Omega_p \rho_{aa} + i\Omega_c \rho_{ba} - 2i\Delta_p \rho_{ca} - \gamma_{ca} \rho_{ca}], \\
\dot{\rho}_{ba} &= \frac{1}{2}[-i\Omega_p \rho_{bc} - 2i\delta \rho_{ba} + i\Omega_c \rho_{ca} - \gamma_{ba} \rho_{ba}].
\end{align}

Next, we solve the above two equations in steady-state, assuming all the atoms in the ground state, we have $\rho_{aa} = 1,\rho_{bb} = 0$, and $\rho_{cc} = 0$.
Under, the weak probe field condition, we drop the $\rho_{bc}$ term in $\dot{\rho}_{ba}$, since it already involves a factor of $\Omega_p$ and remains unpopulated to the lowest order in probe field.
The resulting steady-state solutions are:

\begin{align}
& \rho_{ca} = \frac{i \Omega_p}{\left(\gamma_{ca} + i 2 \Delta_p\right)} + \frac{i \Omega_c}{\left(\gamma_{ca} + i 2 \Delta_p\right)} \rho_{ba}, \\
& \rho_{ba} = \frac{i \Omega_c}{\gamma_{ba} + i 2\delta} \rho_{ca}.
\end{align}

On solving these two coupled equations for $\rho_{ca}$, we obtain the required susceptibility:

\begin{equation}
\begin{aligned}
\chi^{(1)}= & \frac{\left|\mu_{ca}\right|^2 N}{\epsilon_0 \hbar} \\
& \times\left[\frac{4 \delta \left(-\left|\Omega_c\right|^2 + 4 \delta \Delta_p\right) + 4 \Delta_p \gamma_{ba}^2}{\left||\Omega_c\right|^2 + \left(\gamma_{ca} + i 2 \Delta_p\right)\left(\gamma_{ba} + i 2 \delta\right)|^2}\right. \\
& \left. + i \frac{8 \delta^2 \gamma_{ca} + 2 \gamma_{ba} \left(\left|\Omega_c\right|^2 + \gamma_{ba} \gamma_{ca}\right)}{|\left|\Omega_c\right|^2 + \left(\gamma_{ca} + i 2 \Delta_p\right)\left(\gamma_{ba} + i 2 \delta\right)|^2}\right]. 
\end{aligned}
\end{equation}

\newpage
\renewcommand{\bibname}{References}
\bibliography{manuscript}
\end{document}